\documentstyle[12pt,epsf]{article}
\newcommand{\barr}{\begin{array}}
\newcommand{\earr}{\end{array}}
\newcommand{\beq}{\begin{equation}}
\newcommand{\eeq}{\end{equation}}
\newcommand{\bea}{\begin{eqnarray}}
\newcommand{\eea}{\end{eqnarray}}

\newcommand{\prdr}[3]{Phys.\ Rev.\ D~{\bf #1}, #2~(#3)}
\newcommand{\plbr}[3]{Phys.\ Lett.\ {\bf #1},~#2~(#3)}
\newcommand{\npbr}[3]{Nucl.\ Phys.\ {\bf #1},~#2~(#3)}
\newcommand{\rmpr}[3]{Rev.\ Mod.\ Phys. {\bf #1},~#2~(#3)}

\makeatletter

\def\compoundrel#1\over#2{\mathpalette\compoundreL{{#1}\over{#2}}}
\def\compoundreL#1#2{\compoundREL#1#2}
\def\compoundREL#1#2\over#3{\mathrel
	{\vcenter{\hbox{$\m@th\buildrel{#1#2}\over{#1#3}$}}}}
\makeatother

\def\c+{c^{\dagger}}
\def\d+{d^{\dagger}}

\include{psfig}
\begin{document}
\begin{titlepage}
\rightline{\vbox{\halign{&#\hfil\cr
&CUMQ/HEP 97\cr
&\today\cr}}}
\vspace{0.5in}
\begin{center}

{\Huge Bounds on R-Parity Violating Parameters from Fermion EDM's}
\\
\medskip
\vskip0.4in
\normalsize{M.~Frank$^{\sl a,}$\footnote[1] {e-mail mfrank@vax2.concordia.ca}
 and H. Hamidian$^{\sl b,}$\footnote[2]
 {e-mail hamidian@vanosf.physto.se}}
\smallskip
\medskip

{\sl $^{\sl a}$Department of Physics, Concordia University, 1455 De Maisonneuve
Blvd. W.\\ Montreal, Quebec, Canada, H3G 1M8
\\
$^{\sl b}$Department of Physics, Stockholm University, Box 6730, S-113~85\\
Stockholm, 
Sweden}
\end{center}
\vskip1.0in

\begin{center}
{\large\bf Abstract}
\end{center}

\smallskip

We study one-loop contributions to the fermion electric dipole moments
in the Minimal Supersymmetric Standard Model with explicit $R$-parity violating 
interactions. We obtain new individual bounds on $R$-parity violating Yukawa 
couplings and put more stringent limits on certain parameters than those 
obtained 
previously.

\end{titlepage}
\baselineskip=20pt
\newpage
\pagenumbering{arabic}

\noindent{\large\bf Introduction}
\vspace{0.1in}

\noindent The Standard Model (SM) of electroweak interactions is constructed in 
such a way
that it automatically conserves both the baryon number $B$ and the lepton flavor
number $L$. These (accidental) global symmetries of the SM result from the
particle content and the $SU(2)_L \times U(1)_Y$ gauge invariance of the theory
and naturally explain the non-observation of $B$- and $L$-violating processes,
as well as the stability of the nucleon. However, in supersymmetric (SUSY)
extensions of the SM these features no longer follow. In fact, by promoting
the SM fields to superfields, additional gauge- and Lorentz-invariant terms will
be generated which violate $B$ and $L$ conservation. For example, in the SM
the Higgs doublet and the lepton doublet have the same $SU(2) \times U(1)$
quantum numbers, but the spins of the particles are different; whereas in the
SUSY extensions of the SM the distinction between the Higgs doublet and the
lepton doublet disappears and one would naturally expect lepton number
violation. In order to forbid $B$- and $L$-violating interactions in SUSY
extensions of the SM a parity quantum number defined as $R=(-1)^{3B+L+2S}$,
where $S$ is the spin, is assigned to each component field and invariance under
$R$ transformation is imposed \cite{aulakh}. Although the assignment of the {\em
ad hoc} $R$-parity to superfields in the SUSY extensions of the SM reproduces
the global $B$ and $L$ $~U(1)$ symmetries of the SM, it is by no means the only
symmetry which allows the construction of phenomenologically viable SUSY
extensions of the SM \cite{ibanez}. In fact, from a phenomenological point of
view, it is most important to ensure that there are no interaction terms in the
Lagrangian which lead to rapid proton decay and, in this respect, other discrete
symmetries can be used which are even more effective than $R$-parity. This is
simply due the fact that $R$-parity only forbids dimension-four $B$- and
$L$-violating operators in the Lagrangian, while dimension-five operators can
still remain dangerous, even if suppressed with cut-off's as large as the Planck
mass (see Aulakh {\em et al.} in Ref.\cite{aulakh} for an interesting
alternative to $R$-parity which forbids dimension-four and dimension-five
$B$-violating interactions while allowing $L$-violating operators of the same
dimensions in the Lagrangian). Since there is no fundamental theoretical basis 
for 
imposing $R$-parity 
conservation on the minimal supersymmetric extension of the Standard Model 
(MSSM), 
it is 
worthwhile to 
investigate the phenomenological constraints on theoretically allowed 
$R$ parity breaking couplings in the MSSM.

Unlike the SM in which all the elementary matter fields are fermions and, 
consequently, 
$B$ 
and $L$ quantum numbers are separately conserved, the MSSM has a particle 
content 
which 
contains scalar leptons and quarks, thus allowing separate $B$ and $L$ violating 
interaction terms in the Lagrangian.     

Using only the MSSM superfields, the most general renormalizable $R$-parity 
violating 
superpotential can be written as:
\bea
\label{spot1}
W=\lambda_{ijk}L_{i}L_{j}{\bar {E_k}}+\lambda_{ijk}^{'}L_{i}Q_{j}{\bar {D_k}}+
\lambda_{ijk}^{''}{\bar {U_i}}{\bar {D_j}}{\bar {D_k}},
\label{superpot}
\eea
where $i$, $j$, $k$ are generation indices and we have rotated away a term of 
the 
form 
$\mu_{ij}L_{i}H_{j}$. In (\ref{superpot}) $L$ and $Q$ denote the lepton and 
quark 
doublet 
superfields respectively, and ${\bar {U_i}}$, ${\bar {D_j}}$ and ${\bar {D_k}}$ 
are 
singlet 
$SU(2)$ superfields.
The couplings $\lambda_{ijk}$ and 
$\lambda_{ijk}^{''}$ are antisymmetric with respect to the interchange of 
$SU(2)$ 
flavor 
indices: $\lambda_{ijk}=-
\lambda_{jik}$ and $\lambda_{ijk}^{''}=-\lambda_{ikj}^{''}$. 

To avoid rapid proton decay, it is not possible for both
$\lambda$, $\lambda'$ type and $\lambda''$ type couplings to have nonzero 
values. 
We shall 
assume here---as is often done in the literature---that only the lepton number 
is 
violated 
(see \cite{chang} for restrictions on the $\lambda''$ type couplings).

Viewing the SM as a low-energy effective theory, one often searches for 
potential 
contributions arising from the physics beyond the SM. In this manner, numerous 
studies on 
$R$-parity 
violating decays have either resulted in separate bounds on the $\lambda$, 
$\lambda''$ couplings, or on their  
products. Some of the most important studies involve limits coming from proton 
stability, 
$n$--$\bar{n}$ oscillations, $\nu_e$-Majorana mass, neutrinoless double $\beta$
decays, charged current universality, $\nu_\mu$--$e$ deep inelastic scattering, 
atomic parity  violation, $e$--$\mu$--$\tau$ universality, $K^{+}$-decays, 
$\tau$-decays, $D$-decays and precision measurements of LEP electroweak
observables \cite {bhatta}. In addition, when the assumption of $R$-parity 
conservation is 
relaxed, the superpartner spectrum for the MSSM is expected to be dramatically 
different 
from the one with $R$-conservation, the most important consequence being that 
the 
lightest 
supersymmetric particle can decay.

The strongest bounds so far on the $\lambda$ and $\lambda'$ couplings come
 from cosmological considerations on the survival of the cosmic $\Delta B$ 
\cite{bcamp}. They are usually obtained assuming the $L$-violating 
interactions to be constantly out of equilibrium until the weak scale, so that 
they cannot wash out any $(B-L)$ asymmetry previously generated. They are 
relaxed in the case where the $L$-violating interactions are allowed to survive 
at
the weak scale to give rise to lepton-number violations. Lepton-number violating 
interactions in the context of $R$-parity breaking have been studied 
recently and more stingent bounds than those reported earlier have been found 
\cite{huitu mas}.

The most precise low-energy measurements in leptonic physics are the lepton 
flavor-violating decays such as $\mu\rightarrow e\gamma$, 
$\mu\rightarrow eee$, the anomalous magnetic moment of the muon 
$a_{\mu}$, and the EDM of the electron. These are all very
sensitive probes and are often used to explore physics beyond the SM. In 
particular, the electric dipole moment (EDM) of the leptons (especially the 
electron) and 
the neutron are strictly bound by experiments with the currently available upper 
limits given by\cite{Partdata},
\bea
\label{exptedme}
d_{e} && = (3\pm 8)\times 10^{-27} e~cm,
\nonumber\\
d_{\mu} && = (3.7\pm 3.4)\times 10^{-19} e~cm,
\nonumber\\
d_{\tau}&& < (3.7\pm 3.4)\times 10^{-17} e~cm,
\label{leptonedms}
\eea
for the leptons, and
\bea
d_{n}< 1.1\times 10^{-25} e~cm
\label{neutronedm}
\eea
for the neutron. In this letter we investigate bounds on $R$-parity violating 
interactions by 
using the available experimental upper limits on the EDM's of the fermions. As 
we 
shall discuss below, the individual bounds that we put on certain $R$-parity 
violating 
couplings  
are more stringent than those found prior to this work. We shall also briefly 
discuss the 
significance of the EDM bounds compared to the ones obtained by using the---also 
accurately measured---values of the lepton anomalous magnetic moments.   
\\

\noindent{\large\bf $R$-Parity Violating Interactions and Fermion {\rm EDM's}}
\vspace{0.1in}

\noindent  We shall begin with a brief 
review of the fermion 
EDM's and then proceed to evaluate the leading order contributions that arise by 
including $R$-parity violating interactions in the MSSM. 

The electric dipole moment of an elementary fermion is defined through its 
electromagnetic form factor $F_3(q^2)$ found from the (current) matrix element
\bea
\label{formfactors}
\langle f(p')|J_{\mu}(0)|f(p) \rangle=\bar{u}(p')\Gamma_{\mu}(q)u(p),
\eea
where $q=p'-p$ and
\bea
\label{current}
\Gamma_{\mu}(q)=F_1(q^2)\gamma_{\mu}+F_2(q^2)i\sigma_{\mu\nu}q^{\nu}/2m 
+F_A(q^2)
(\gamma_{\mu}\gamma_5q^2-2m\gamma_5q_{\mu})+F_3(q^2)\sigma_{\mu\nu}
\gamma_5q^{\nu}/2m,
\eea
with $m$ the mass of the fermion. The EDM of the fermion field $f$ is then given 
by
\bea
\label{edm}
d_f=-F_3(0)/2m,
\eea
corresponding to the effective dipole interaction
\bea
\label{dipole}
{\cal L}_I= - \frac{i}{2} d_f \bar{f}\sigma_{\mu\nu}\gamma_5 f F^{\mu\nu}
\eea
in the static limit.

Since a non-vanishing $d_f$ in the SM results in fermion chirality flip, it 
requires both 
$CP$ violation and $SU(2)_L$ symmetry breaking. Even if one allows for 
$CP$-violation in the leptonic 
sector of the SM, the lepton EDM's vanish to one-loop order due to the 
cancellation of all the 
$CP$-violating phases. Two-loop calculations for the electron \cite{donoghue} 
and for quarks 
\cite{shabalin} also yield a zero EDM. In the MSSM, however, there are many more 
sources of $CP$ 
violation 
than in the SM. In addition to the usual Kobayashi-Maskawa phase 
$\delta$ from the quark mixing matrix, there are phases arising from complex 
parameters in the superpotential and in the soft supersymmetry breaking terms. 
The phases of particular interest to us are those coming from the so-called 
$A$-terms, $A_{u,d} = |A_{u,d}|~\exp ~(i\phi_{A_u,d})$ ~\cite{dugan}.  The 
$CP$-violating 
effects arise from the squark mass matrix which has the following form:
\bea
\label{squarmassmatr}
{\cal L}_{M_{\tilde u}}=(\tilde{u}_L^\dagger\;
\tilde{u}_R^\dagger)
\left(
\begin{array}{cc}
\mu_L^2+m_u ^2& A_u^\ast m_u\\
 A_um_u &\mu_R^2+m_u^2 ,
\end{array}
\right)
\left(\begin{array}{c}
\tilde{u}_L\\
\tilde{u}_R
\end{array}\right),
\label{slepmassmatr}
\eea
and similarly for ${\cal L}_{M_{\tilde d}}$, where the mass parameters 
$|A_u|$, $\mu_L$ and $\mu_R$ are expected to be of the order
of the $W$-boson mass $M_W$. The fields ${\tilde u}_L$, ${\tilde u}_R$ can be 
transformed 
into mass eigenstates ${\tilde u}_1$, ${\tilde u}_2$,
\bea
{\tilde u}_L &=& \exp (-\frac{1}{2}i \phi_{A_u})(\cos \theta~ {\tilde u}_1 + 
\sin
\theta ~{\tilde u}_2),\nonumber \\
{\tilde u}_R &=& \exp (\frac{1}{2}i \phi_{A_u})(\cos \theta ~{\tilde u}_2 - \sin
\theta ~{\tilde u}_1),
\label{masseig}
\eea
where the mixing angle $\theta$ is given by
\bea
\label{theta}
\tan 2\theta=2|A_u|m_u/(\mu_L^2-\mu_R^2).
\eea
and the physical masses, $M_{1,2}$, corresponding to the eigenvalues of the mass 
matrix in 
(\ref{slepmassmatr}) are
\bea
M_{1,2}^2 = \frac{1}{2}{\mu_L^2+\mu_R^2+2m_u^2 \pm [(\mu_L^2-\mu_R^2)^2 + 
4m_u^2 |A_u|^2]^{1/2}}.
\label{eigen}
\eea

The lepton EDM's at one-loop order are generated by the interactions in Fig.1 
(with similar Feynman diagrams for the muon and the tau) and resemble those in 
the MSSM with charginos or neutralinos in the loop. The 
contributions to the 
lepton EDM's are then given by
\\
\bea
d_{e_i} =&-&|\lambda'_{ijk}|^2 \frac{4e}{3} \frac{m_{d_k}}{m_{\tilde 
f}^3}|A_{u_j}|
\sin\theta\cos\theta\sin(\phi_{A_u}) f_3(x_{d_k})\nonumber\\
&-&|\lambda'_{ijk}|^2 \frac{2e}{3} \frac{m_{d_k}}{m_{\tilde f}^3}|A_{u_j}|
\sin\theta\cos\theta\sin(\phi_{A_u}) f_4(x_{d_k})\nonumber\\
&-&|\lambda'_{ijk}|^2 \frac{2e}{3} \frac{m_{u_j}}{m_{\tilde f}^3}|A_{d_k}|
\sin\theta\cos\theta\sin(\phi_{A_d}) f_3(x_{u_j})\nonumber\\
&-&|\lambda'_{ijk}|^2 \frac{4e}{3} \frac{m_{u_j}}{m_{\tilde f}^3}|A_{d_k}|
\sin\theta\cos\theta\sin(\phi_{A_d}) f_4(x_{u_j}),
\label{edm1}
\eea
where $x_{u,d}=(m_{u,d}/m_{\tilde f})^2$, with ${\tilde f}$ the scalar 
quark in the loop, and the loop integrals are expressed in 
a familiar form in terms of the functions

\bea
f_3(x) &=& \frac {1}{2(1-x)^2}\left[ 1+x+\frac{2x\ln x}{1-x} \right]\nonumber\\
f_4(x) &=& \frac{1}{2(1-x)^2}\left[ 3-x+\frac {2 \ln x}{1-x} \right].
\label{loopint}
\eea 
\\
\noindent In order to simplify, we will assume degenerate squark masses,   
$\mu_L \approx \mu_R \approx |A_{u,d}| = {\cal O}(M_W)$, and expand only to the 
leading order in 
$m_{u,d}|A_{u,d}|/M_{1,2}^2$. Comparing the above expressions for the fermion 
EDM's with 
the usual ones obtained in the 
MSSM \cite {kizuk}, it is not difficult to see that there are 
two sources of enhancement: one coming from the absence of the electroweak 
coupling constant, $\alpha_{ew}$,  responsible for an enhancement of ${\cal 
O}(10^2)$, and 
another 
from potentially large fermionic masses in the 
loop. Indeed, in this scenario it is possible to obtain a contribution to the 
electron EDM 
proportional to the mass of the top quark, in contrast to the usual one 
proportional to 
$m_e$. (Note that even if the EDM is proportional to the mass of the up quark, 
there will still be an enhancement of ${\cal O}(20)$.) Unfortunately estimating 
$d_e$, $d_{\mu}$, and  $d_{\tau}$  numerically is not 
completely 
straightforward since no model-independent experimental information is available 
on squark 
masses and mixing angles. We shall assume, without great loss of 
generality, that $\cos \theta = \sin \theta =1/\sqrt{2}$.  We shall also assume,
 $\phi_{A_u} =\phi_{A_d}$ and $|A_{u_j,d_k}|\approx m_{\tilde f}= {\cal 
O}(M_W)$, 
which is in 
agreement with the naturalness of the MSSM. Putting all these together, Eq. 
(\ref{edm1}) 
becomes
\\
\bea
d_{e_i}=&-&|\lambda'_{ijk}|^2 \frac{1}{3} (\frac{m_{\tilde f}}
         {100 GeV})^{-3} (\frac{|A_{u,d}|}{100 GeV}) \sin (\phi_A) \nonumber\\
         &\times& \left\{ m_{d_k} \left[ 4f_3(x_{u_j}) + 2f_4(x_{u_j}) \right] +
	m_{u_j} \left[ 2f_3(x_{d_k}) + 4f_4(x_{d_k}) \right] \right\} 
	\times 10^{-21}~e~cm.
\label{edm2}
\eea
\\
We would like to comment that in addition to these contributions, in theories 
with 
massive 
neutrinos one could have contributions coming from the Feynman diagrams such as 
those in 
Fig.2. 
These  
are not included here since we restrict ourselves to the particle spectrum of 
the MSSM. Including these contributions in other SUSY extensions of the SM could 
provide 
restrictions on the $\lambda_{ijk}$ parameters, however they would all depend 
rather 
sensitively on the neutrino mass.  

Any estimate of the lepton EDM's must be correlated and 
further restricted by estimates of the neutron EDM. In the $R$-parity-conserving 
MSSM the 
neutron EDM severely restricts the masses of the squarks, barring accidental 
cancellations  
between the supersymmetric phases in the squark and gluino matrices. With the 
introduction of $R$-parity violating interactions, the terms contributing to the 
quark EDM's are Yukawa-type only, such as those shown in Fig.3 for the up quark. 
Taking all the scalar quark masses to be the 
same, and taking $m_u \simeq m_d =10~MeV$, the up- and down- quark 
EDM's are equal and the neutron EDM is $d_n=\frac{4}{3}~ d_d - \frac{1}
{3}~d_u \simeq d_d$. Since the experimental limit on the neutron EDM is
$|d_n|< 1.2\times 10^{-25} e~cm$ the limits obtained on the 
$\lambda_{ijk}$ parameters are weaker than those obtained from the electron EDM,
in contrast with the existing situation in $R$-parity conserving MSSM.

The limits that we obtain on the individual $\lambda'_{ijk}$ parameters 
from the electron ($\lambda'_{1jk}$), muon ($\lambda'_{2jk}$)   
EDM's are given in Table~1. Unfortunately, the weak 
constraint on the tau EDM is insufficient to adequately restrict all the 
$\lambda'_{3jk}$
coeficients. Two cases are considered in Table 1: (i) light slepton spectrum 
$(m_{\tilde f} 
= 100~GeV)$ and (ii) heavy slepton spectrum $(m_{\tilde f} =1~TeV)$. In both 
cases 
we find 
strong bounds from the electron EDM and weaker 
ones from the $\mu$ or $\tau$ EDM's. The strongest limits  
appear to be the ones on $\lambda_{1j3}$ because of  effects of ${\cal O}(m_t)$.
The difference between a light and a heavy squark spectrum is in agreement
with other estimates on the $\lambda'$ parameters. The different scenarios 
bridge the gap of any other possible assumptions on the superparticle spectrum. 
For instance, if instead of assuming a degenerate squark mass spectrum, we 
assume 
a 
universal
scalar mass spectrum at the GUT scale and then evolve masses at the 
electroweak scale as in \cite{ramond}, our assumptions resemble very closely 
the case in which $\tan\beta \leq 10$ . In this case the squarks of the first 
two families and $\tilde {b}_R$ have very similar masses and only the stop is 
heavier. That would affect mostly the $\lambda_{i3k}$ coefficients;   
 the results will be of the same
order of magnitude with slightly different numerical factors.

If we assume, as has been sometimes done in the literature, that the 
$\lambda'_{ijk}$ 
are flavor blind, i.e. $\lambda'_{ijk}= \lambda'$, we will be able 
to severely restrict all the $R$-parity violating couplings, as shown in Figures 
4 and 5 respectively for the light and heavy squark mass scenarios. In both 
cases the restrictions 
on 
the 
$\lambda'_{ijk}$ couplings are more stringent than previously found limits. 
The EDM's do not, unfortunately, provide any limits on the $\lambda_{ijk}$ or 
$\lambda''_{ijk}$ 
 couplings, so one would have to rely on limits obtained by considering other
phenomena, such as those mentioned in the introduction.

We shall now briefly comment on possible restrictions coming from the anomalous 
magnetic moment of the muon, $a_{\mu}$. Its experimental value is 
$a_{\mu}^{exp}=
1165922(9) \times 10^{-9}$ ~\cite {Partdata} and its measured deviation from 
the SM prediction lies within a range of $-2 \times 10^{-8} \leq 
\Delta a_{\mu}^{exp} \leq 2.6 \times 10^{-8}$. The one-loop contributions  
to the muon magnetic moment also come from Feynman diagrams similar to those in 
Fig.1 and are 
given 
by
\bea
\label{g-2}
a_{\mu} = F_2(0)/e,
\eea
where $F_2(0)$ is the static limit of the electromagnetic form factor defined in 
(\ref 
{current}). Unfortunately, the restrictions that $a_{\mu}$ would place on the 
$R$-parity 
violating couplings $\lambda$ and $\lambda'$
are extremely weak compared to those coming from the EDM's. 
The EDM of the electron is expected to be known to 
about five orders of magnitude more accurately than its anomalous
magnetic moment within a few years. Indeed the precise measurements of the 
electron $(g-2)$ factor yield $\Delta[(g-2)/2]=1\times 10^{-11}$, which 
corresponds to 
$\Delta(F_2(0)/2m_e)= 2\times 10^{-22} e~cm$. The same is true for the anomalous 
magnetic moment of the 
muon \cite{bern}. In addition, for the calculation of the form factor 
$F_2(q^2)$ for the electron (muon) the helicity flip 
occurs on the external fermion line, giving rise to an amplitude proportional 
to the electron (muon) mass. These two factors combined, give weaker bounds on 
the 
$R$-parity violating Yukawa couplings than 
previously found, despite the fact that the anomalous magnetic 
moment does not require $CP$ violation.
  
To conclude, we have studied one-loop contributions to the lepton and neutron 
EDM's in the 
MSSM with explicit $R$-parity violating interactions. We have used the---very 
accurately 
measured---experimental values of the lepton EDM's to severely restrict the 
magnitude of 
the $R$-parity violating Yukawa couplings. This analysis has allowed us to 
obtain 
new bounds on 
individual $\lambda'_{ijk}$ couplings, rather than their combinations with other 
parameters.  

\begin{center}
{\bf Acknowledgement}
\end{center}

This work was supported by NSERC and the Swedish Natural Science Research 
Council. 
M.F. 
would like to thank the Field Theory and Particle Physics Group of Stockholm 
University 
(where this work was initiated) for their hospitality.

\bibliographystyle{plain}

\newpage

\begin{center}
{\huge Figure Captions:}
\end{center}

\vskip0.2in

\noindent {\bf Figure 1}: Graphs contributing to the electron EDM that put 
limits 
on the
value of the $\lambda'_{1jk}$ couplings. Similar graphs contribute to the $\mu$- 
and $\tau$-EDM's and put limits on $\lambda'_{2jk}$ and $\lambda'_{3jk}$ 
respectively.

\vskip0.2in

\noindent {\bf Figure 2}: Graphs that could contribute to the electron EDM in 
theories with massive 
neutrinos. Similar graphs contribute to the $\mu$- and $\tau$-EDM's. These 
graphs could restrict the values of the $\lambda_{ijk}$ couplings.

\vskip0.2in

\noindent {\bf Figure 3}: Graphs contributing to up quark EDM's and therefore to 
the 
neutron EDM. For the down quark the same graphs contribute, with 
$d \leftrightarrow u$ and $e \leftrightarrow e^c$.    

\vskip0.2in

\noindent {\bf Figure 4}: The electron $EDM$ as a function of a universal 
coupling 
$\lambda'_{ijk}=\lambda'$ (horizontal axis) for the light squark scenario , 
$m_{\tilde f}=100~GeV$. We take the following values for the quark masses 
masses:
$m_u=m_d=10~MeV$ , $m_s=300~MeV$ , $m_c=1.5~GeV$, $m_b=4.5~GeV$ and 
$m_t=175~GeV$ \cite {Partdata}.

\vskip0.2in

\noindent {\bf Figure 5}: The electron $EDM$ as a function of a universal 
coupling 
$\lambda'_{ijk}=\lambda'$ (horizontal axis) for the  
heavy squark scenario , 
$m_{\tilde f}=1~TeV$. We take the following values for the quark masses masses:
$m_u=m_d=10~MeV$ , $m_s=300~MeV$ , $m_c=1.5~GeV$, $m_b=4.5~GeV$ and 
$m_t=175~GeV$
 ~\cite {Partdata}.
\\
\\
\\

\begin{center}
{\huge Table Captions:}
\end{center}

\vskip0.2in

\noindent {\bf Table 1}: Bounds on the $R$-parity violating parameters, 
$\lambda_{ijk}'$, from the 
electron ($\lambda_{1jk}'$) and muon ($\lambda_{2jk}'$) EDM's, compared with 
previous bounds obtained from: (a) $K^+$-decay \cite{agashe}; (b) Atomic 
parity violation and $eD$ asymmetry \cite{barger}; (c) $t$-decay 
\cite{agashe}; (d) 
$\nu_e$-Majorana mass \cite{godbole}; (e) $\nu_{\mu}$ deep-inelastic scattering 
\cite{barger}.

\newpage  
\begin{center}
{\bf Table 1}
\vskip0.2in
\begin{tabular}{|lrcr|} 
\multicolumn{1}{c}{$|\lambda'_{ijk}|^2 \leq $} &
\multicolumn{1}{c}{$m_{\tilde f}=100~GeV$} &
\multicolumn{1}{c}{$m_{\tilde f}=1~TeV$} &
\multicolumn{1}{c}{Previous Limits for $m_{\tilde f}=100~GeV$}
\\
\hline
$|\lambda'_{111}|^2$      &  $3 \times 10^{-9}$     &  $2.4 \times 10^{-7}$ 
				&   $1.44 \times 10^{-4}$ (a)  \\
$|\lambda'_{112}|^2$      &  $5 \times 10^{-10}$    &  $3 \times 10^{-8}$ 
				&   $1.44 \times 10^{-4}$ \\                       
$|\lambda'_{113}|^2$      &  $9 \times 10^{-11}$    &  $4 \times 10^{-9}$ 
				&   $1.44 \times 10^{-4}$ (a) \\  
$|\lambda'_{121}|^2$      &  $7.5 \times 10^{-11}$    &  $4.3 \times 10^{-9}$ 
				&   $1.44 \times 10^{-4}$ (a) \\  
$|\lambda'_{122}|^2$      &  $6.6 \times 10^{-11}$    &  $4 \times 10^{-9}$ 
				&   $1.44 \times 10^{-4}$ (a) \\  
$|\lambda'_{123}|^2$      &  $4 \times 10^{-11}$    &  $2 \times 10^{-9}$ 
				&   $1.44 \times 10^{-4}$ (a) \\  
$|\lambda'_{131}|^2$      &  $2.6 \times 10^{-12}$  &  $2.4 \times 10^{-11}$ 
				&   $1.44 \times 10^{-4}$ (a), $0.0676$ 
(b) \\  
$|\lambda'_{132}|^2$      &  $2.6 \times 10^{-12}$  &  $2.4 \times 10^{-11}$ 
				&   $1.44 \times 10^{-4}$ (a), $0.16$ (c) 
\\  
$|\lambda'_{133}|^2$      &  $2.5 \times 10^{-12}$  &  $2.3 \times 10^{-11}$ 
				&   $1.44 \times 10^{-4}$ (a), $10^{-6}$ 
(d) \\  
\hline
$|\lambda'_{211}|^2$      &  $3 \times 10^{-1}$     &  $24$ 
				&   $1.44 \times 10^{-4}$ (a) \\  
$|\lambda'_{212}|^2$      &  $5 \times 10^{-2}$     &  $3$ 
				&   $1.44 \times 10^{-4}$ (a) \\                         
$|\lambda'_{213}|^2$      &  $9 \times 10^{-3}$     &  $4 \times 10^{-1}$ 
				&   $1.44 \times 10^{-4}$ (a) \\  
$|\lambda'_{221}|^2$      &  $7.5 \times 10^{-3}$     &  $4.3 \times 10^{-1}$ 
				&   $1.44 \times 10^{-4}$ (a) \\  
$|\lambda'_{222}|^2$      &  $6.6 \times 10^{-3}$     &  $4 \times 10^{-1}$ 
				&   $1.44 \times 10^{-4}$ (a) \\  
$|\lambda'_{223}|^2$      &  $4 \times 10^{-3}$     &  $2 \times 10^{-1}$ 
				&   $1.44 \times 10^{-4}$ (a) \\  
$|\lambda'_{231}|^2$      &  $2.6 \times 10^{-4}$   &  $2.4 \times 10^{-3}$ 
				&   $1.44 \times 10^{-4}$ (a), $0.0484$ 
(e) \\  
$|\lambda'_{232}|^2$      &  $2.6 \times 10^{-4}$   &  $2.4 \times 10^{-3}$ 
				&   $1.44 \times 10^{-4}$ (a), $0.16$ (c) 
\\  
$|\lambda'_{233}|^2$      &  $2.5 \times 10^{-4}$   &  $2.3 \times 10^{-3}$ 
				&   $1.44 \times 10^{-4}$ (a), $0.16$ (c) 
\\  
\hline

\end{tabular}
\end{center}


\newpage
\thispagestyle{empty}
\begin{figure}[hbtp]
\centerline{\epsffile{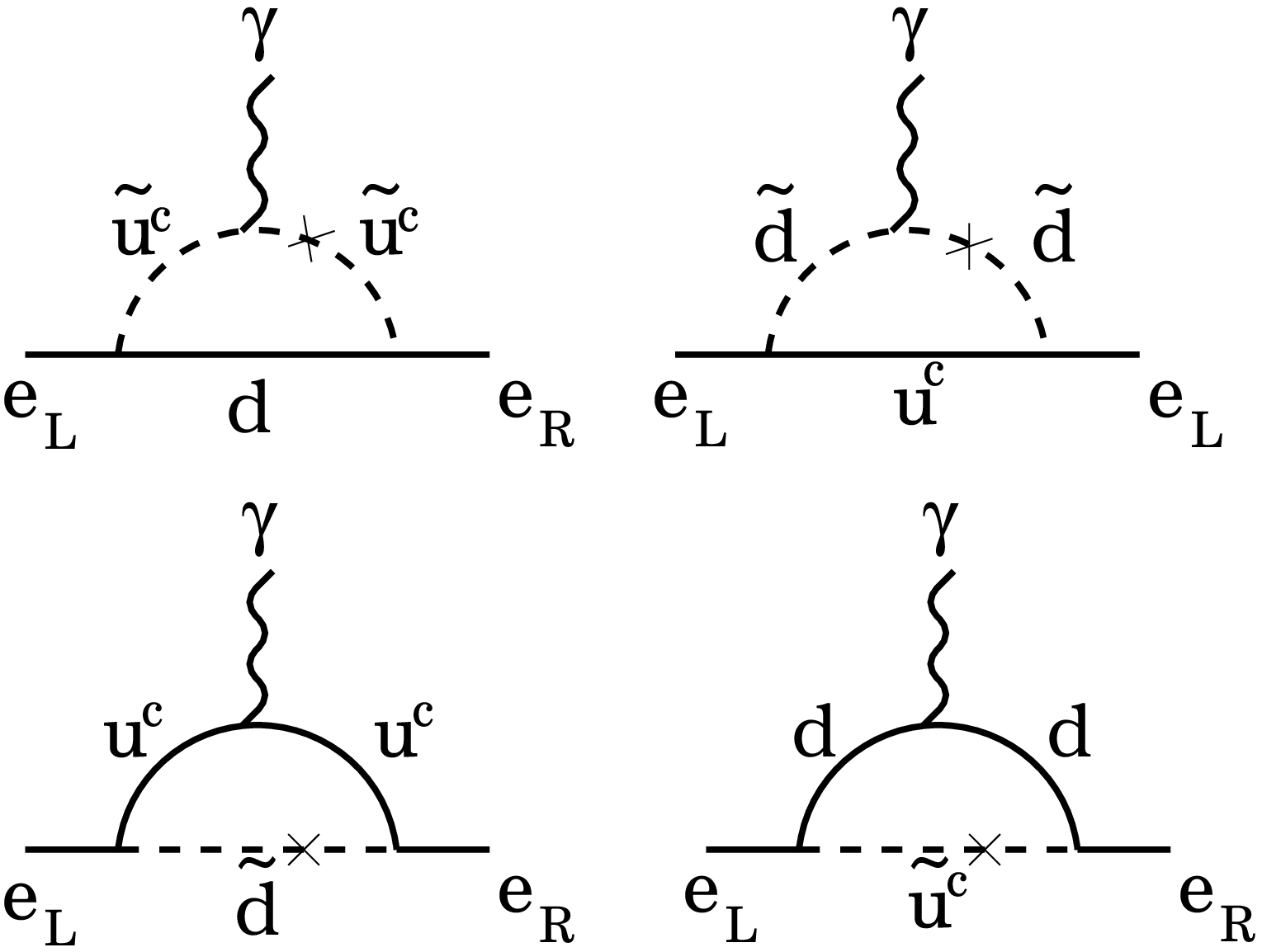}}
\end{figure}
\begin{center}
{\Large Figure 1}
\end{center}

\newpage
\thispagestyle{empty}
\begin{figure}[hbtp]
\centerline{\epsffile{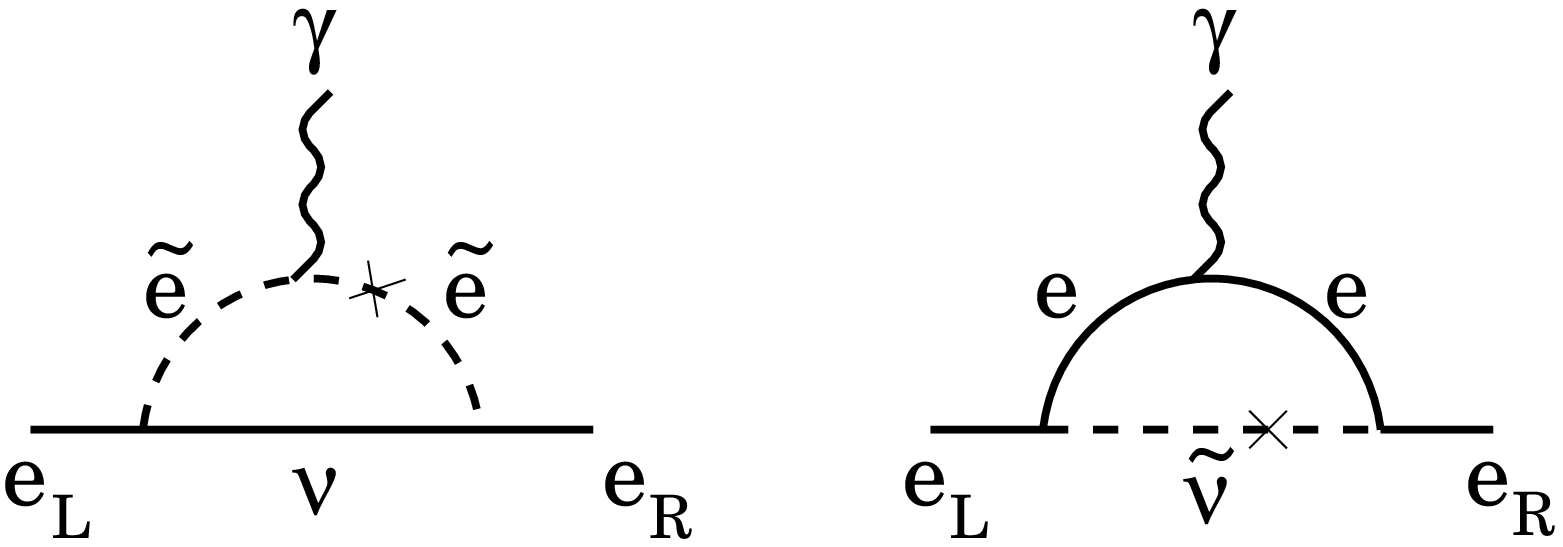}}
\end{figure}
\begin{center}
{\Large Figure 2}
\end{center}

\newpage
\thispagestyle{empty}
\begin{figure}[hbtp]
\centerline{\epsffile{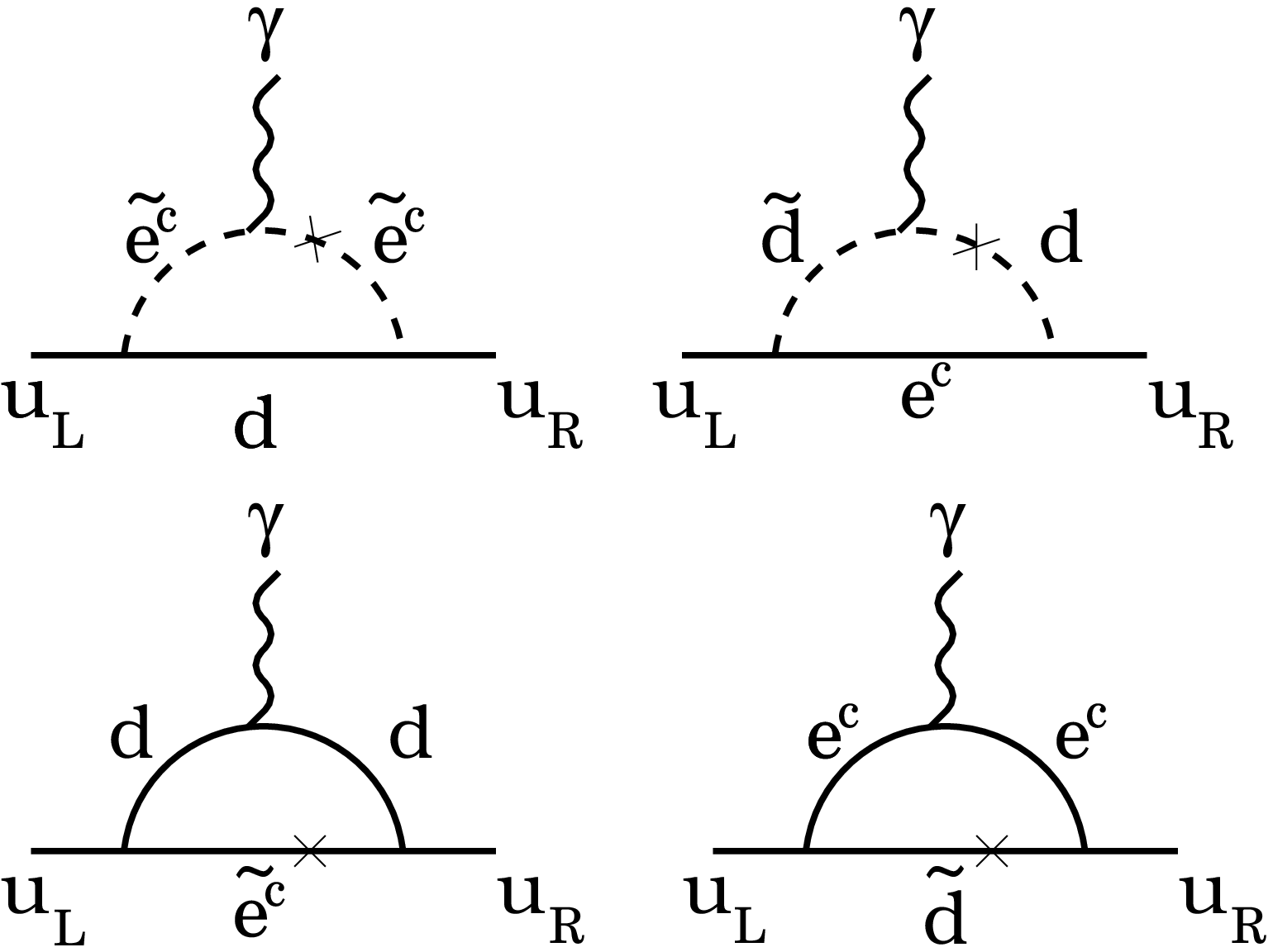}}
\end{figure}
\begin{center}
{\Large Figure 3}
\end{center}


\newpage
\pagestyle{empty}
\begin{figure}[hbtp]
\centerline{\epsffile{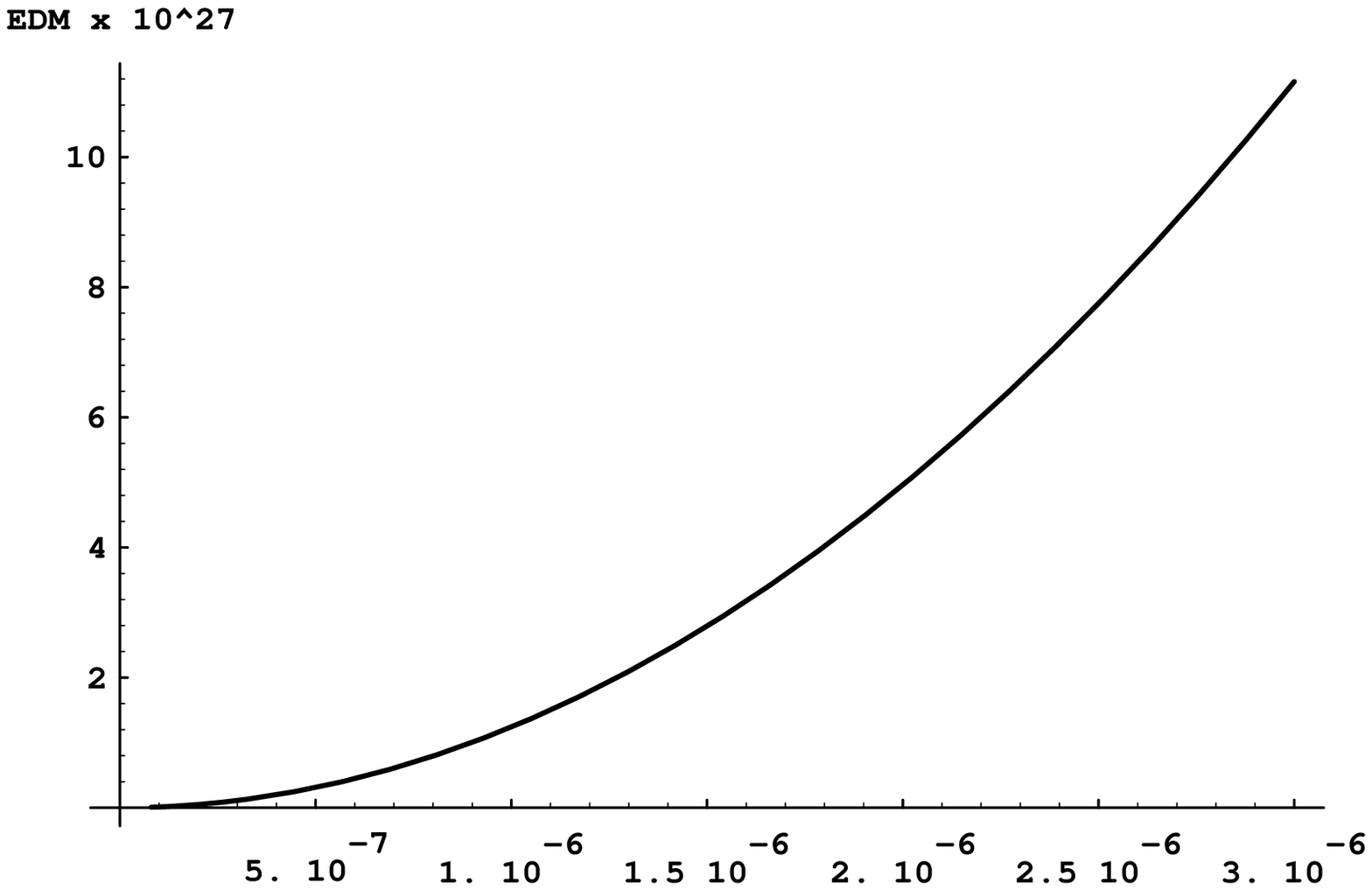}}
\begin{center}
{\Large Figure 4}
\end{center}
\end{figure}

\newpage
\pagestyle{empty}
\begin{figure}[hbtp]
\centerline{\epsffile{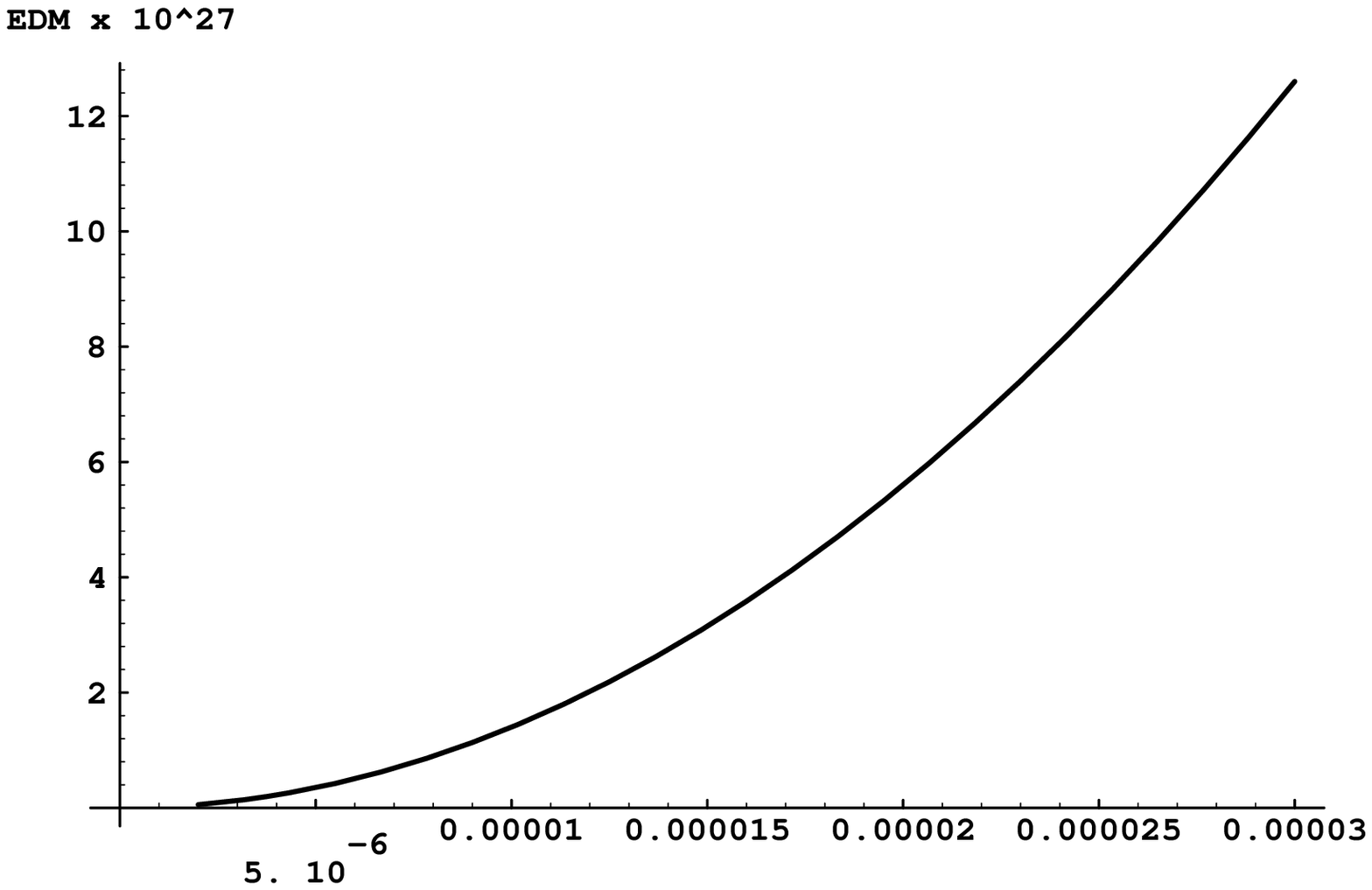}}
\begin{center}
{\Large Figure 5}
\end{center}
\end{figure}


\end{document}